\documentclass{birkjour}
\usepackage{amsmath, amsthm, amssymb, amscd, mathrsfs, eucal,
  amsfonts, latexsym}
 \newtheorem{thm}{Theorem}[section]

 \theoremstyle{definition}
 \newtheorem{defn}[thm]{Definition}
 \theoremstyle{remark}

 \numberwithin{equation}{section}

\begin{document}

%
%
%
%
%
%
%
%
%

\title[Local Thermal Equilibrium and Cosmology]
 {Local Covariance, Renormalization Ambiguity, and Local Thermal Equilibrium  
  in Cosmology}

\author[Rainer Verch]{Rainer Verch}

\address{%
Institut f\"ur Theoretische Physik\\
Universit\"at Leipzig\\
Vor dem Hospitaltore 1\\
D-04103 Leipzig \\
Germany}

\email{verch@itp.uni-leipzig.de}




\begin{abstract}
This article reviews some aspects of local covariance and of the ambiguities
and anomalies involved in the definition of the stress energy tensor of 
quantum field theory in curved spacetime. Then, a summary is given of 
the approach proposed by Buchholz et al.\ to define local thermal equilibrium
states in quantum field theory, i.e., non-equilibrium states to which, locally,
one can assign thermal parameters, such as temperature or thermal stress-energy.
The extension of that concept to curved spacetime is discussed and some
related results are presented. Finally, the recent approach to cosmology by
Dappiaggi, Fredenhagen and Pinamonti, based on a distinguished fixing
of the stress-energy renormalization ambiguity in the setting of the
semiclassical Einstein equations, is briefly described. The concept of
local thermal equilibrium states is then applied, to yield the result that 
the temperature behaviour of a quantized, massless, conformally coupled
linear scalar field at early cosmological times is more singular than that
of classical radiation.   
\end{abstract}

\maketitle

\section{Local Covariant Quantum Field Theory}

The main theme of this contribution is the concept of local thermal
equilibrium states and their properties in quantum field theory on
cosmological spacetimes. The starting point for our considerations
is the notion of local covariant quatum field theory which has been 
developed in \cite{VerSPST, BFV, HolWald} and which is also reviewed
in \cite{BGP} and is, furthermore, discussed in Chris Fewster's
contribution to these conference proceedings \cite{cjfgm}. I will therefore attempt
to be as much as possible consistent with Fewster's notation and 
shall at several points refer to his contribution for precise definitions
and further discussion of matters related to local covariant quantum field
theory.

The basic idea of local covariant quantum field theory is to consider
not just a quantum field on some fixed spacetime, but simultaneously
the ``same'' type of quantum field theory on all --- sufficiently nice ---
spacetimes at once. In making this more precise, one collects all spacetimes
which one wants to consider in a category {\sf Loc}. By definition, an
object in {\sf Loc} is a four-dimensional, globally hyperbolic spacetime
with chosen orientation and time-orientation. An arrow, or morphism of
$\boldsymbol{M} \overset{\psi}{\longrightarrow} \boldsymbol{N}$ of
{\sf Loc} is an isometric, hyperbolic embedding which preserves orientation
and time-orientation. (See Fewster's contribution for more details.)

Additionally, we also assume that we have, for each object $\boldsymbol{M}$ of {\sf Loc},
a quantum field $\Phi_{\boldsymbol{M}}$, thought of as describing some physics
happening in $\boldsymbol{M}$.  More precisely, we assume that there is
a topological $*$-algebra $\mathscr{A}(\boldsymbol{M})$ (with unit) and that
$f \mapsto \Phi_{\boldsymbol{M}}(f)$, $f \in C_0^\infty(\boldsymbol{M})$, is an
operator-valued distribution taking values in $\mathscr{A}(\boldsymbol{M})$.
(This would correspond to a ``quantized scalar field'' which we treat here for
simplicity, but everything can be generalized to more general tensor- and
spinor-type fields, as e.g.\ in \cite{VerSPST}.) One would usually require that 
the $\Phi_{\boldsymbol{M}}(f)$ generate $\mathscr{A}(\boldsymbol{M})$
in a suitable sense, if necessary allowing suitable completions. The unital
topological $*$-algebras also form a category which shall be denoted by {\sf Alg},
where the morphisms 
$\mathscr{A} \overset{\alpha}{\longrightarrow} \mathscr{B}$ are
injective, unital, topological $*$-algebraic morphisms between the objects. 
Then one says that family the $(\Phi_{\boldsymbol{M}})$, as $\boldsymbol{M}$ ranges
over the objects of {\sf Loc}, is a {\it local covariant quantum field theory} if
the assignments $\boldsymbol{M} \to \mathscr{A}(\boldsymbol{M})$ 
and $\mathscr{A}(\psi)(\Phi_{\boldsymbol{M}}(f)) = \Phi_{\boldsymbol{N}}(\psi_*f)$,
for any morphism
$\boldsymbol{M} \overset{\psi}{\longrightarrow} \boldsymbol{N}$ of
{\sf Loc},
induce a functor $\mathscr{A} : {\sf Loc} \to {\sf Alg}$. (See again Fewster's
contribution \cite{cjfgm} for a fuller dish.) Actually, a generally covariant quantum field theory
should more appropriately be viewed as a natural transformation and we refer
to \cite{BFV} for a discussion of that point. The degree to which the present
notion of local covariant quantum field theory describes the ``same'' quantum field
on all spacetimes is analyzed in Fewster's contribution to these proceedings.

We write $\boldsymbol{\Phi} = (\Phi_{\boldsymbol{M}})$ to denote a local
covariant quantum field theory, and we remark that examples known
so far include the scalar field, the Dirac field, the Proca field, and --- with some
restrictions --- the electromagnetic field, as well as perturbatively constructed
$P(\phi)$ and Yang-Mills models \cite{HolWald,HolYM,VerSPST}.
 A {\it state} of a local covariant quantum
field theory $\boldsymbol{\Phi}$ is defined as a family $\boldsymbol{\omega} =
(\omega_{\boldsymbol{M}})$, $\boldsymbol{M}$ ranging over the objects of 
{\sf Loc}, where each $\omega_{\boldsymbol{M}}$ is a state on 
$\mathscr{A}(\boldsymbol{M})$ --- i.e. an expectation value functional. It might
appear natural to assume that there is an invariant state $\boldsymbol{\omega}$,
defined by 
\begin{equation} \label{invast}
 \omega_{\boldsymbol{N}} \circ \mathscr{A}(\psi) = \omega_{\boldsymbol{M}}
\end{equation}
for all morphisms $\boldsymbol{M} \overset{\psi}{\longrightarrow} \boldsymbol{N}$
of {\sf Loc}, but it has been shown that this property is in conflict with the 
dynamics and stability properties one would demand of the $\omega_{\boldsymbol{M}}$
for each $\boldsymbol{M}$ \cite{BFV,HolWald}; the outline of an argument against
an invariant state with the property \eqref{invast} is given in Fewster's contribution
\cite{cjfgm}. 

\section{The Quantized Stress Energy Tensor}


A general property of quantum field theories is the 
existence of a causal dynamical law, or time-slice axiom. In a local
covariant quantum field theory $\boldsymbol{\Phi}$, this leads to a covariant dynamics \cite{BFV}
termed ``relative Cauchy evolution'' in Fewster's contribution \cite{cjfgm}, to which
we refer again for further details. The relative Cauchy evolution consists,
for each $\boldsymbol{M}$ in {\sf Loc}, of an isomorphism
$$ {\rm rce}_{\boldsymbol{M}}(h) : \mathscr{A}(\boldsymbol{M}) \to
 \mathscr{A}(\boldsymbol{M}) $$
which describes the effect of an additive perturbation of the metric
of $\boldsymbol{M}$ by a symmetric 2-tensor field $h$ on the propagation
of the quantum field, akin to a scattering transformation brought about 
by perturbation of the background metric. One may consider the derivation --
assuming it exists -- which is obtained as 
$\left. d/ds \right|_{s=0}{\rm rce}_{\boldsymbol{M}}(h(s))$ where $h(s)$ is
any smooth family of metric perturbations with $h(s=0) = 0$. Following
the spirit of Bogoliubov's formula \cite{Bogo}, this gives rise (under fairly general
assumptions) to a local covariant quantum field $({\sf T}_M)$ which
generates the derivation upon taking commutators, and can be identified
(up to some multiplicative constant) with the quantized stress energy tensor
(and has been shown in examples to agree indeed with the derivation induced
by forming the commutator with the quantized stress energy tensor \cite{BFV}).
Assuming for the moment that ${\sf T}_{\boldsymbol{M}}(f)$ takes 
values in $\mathscr{A}(\boldsymbol{M})$,
we see that in a local covariant quantum field theory which fulfills the time-slice
axiom the stress energy tensor is characterized by the following properties:
\\[6pt]
{\bf LCSE-1} \quad {\sl Generating property for the relative Cauchy evolution}:
\begin{equation}
 2i [{\sl T}_{\boldsymbol{M}}(f),A] = \left. \frac{d}{ds}\right|_{s = 0} {\rm rce}_{\boldsymbol{M}}(h(s))
(A) \,, \quad A \in \mathscr{A}(\boldsymbol{M})\,, \quad f = \left.\frac{d}{ds}\right|_{s=0} h(s)
\end{equation}
Note that $f = f^{ab}$ is a    
smooth, symmetric $C_0^\infty$ two-tensor field on $\boldsymbol{M}$. Using
the more ornamental abstract index notation, one would therefore write
$${\sf T}_{\boldsymbol{M}ab}(f^{ab}) = {\sf T}_{\boldsymbol{M}}(f) \,.$$
{\bf LCSE-2} \quad {\sl Local covariance}:
\begin{equation}
\mathscr{A}(\psi) \left({\sf T}_{\boldsymbol{M}}(f) \right) = 
 {\sf T}_{\boldsymbol{N}}(\psi_*f)
\end{equation}
whenever $\boldsymbol{M} \overset{\psi}{\longrightarrow} \boldsymbol{N}$
is an arrow in {\sf Loc}.
\\[6pt]
{\bf LCSE-3} \quad {\sl Symmetry}:
\begin{equation}
 {\sf T}_{\boldsymbol{M}ab} = {\sf T}_{\boldsymbol{M}ba} \,.
\end{equation}
{\bf LCSE-4} \quad {\sl Vanishing divergence}:
\begin{equation}
 \nabla^a{\sf T}_{\boldsymbol{M}ab} = 0\,,
\end{equation}
where $\nabla$ is the covariant derivative of the metric of $\boldsymbol{M}$.
\\[6pt]
Conditions {\bf LCSE-3,4}
are (not strictly, but morally) consequences of the previous two conditions,
see \cite{BFV} for discussion.
 
The conditions {\bf LCSE-1...4} ought to be taken
as characterizing for any stress energy tensor of a given local covariant 
quantum field theory $\boldsymbol{\Phi}$.
However, they don't fix 
${\sf T}_{\boldsymbol{M}}$ completely. Suppose that  we have made some
choice, $({\sf T}^{[1]}_{\boldsymbol{M}})$, of a local covariant stress energy
tensor for our local covariant quantum field theory $\boldsymbol{\Phi}$. Now
choose a local covariant family $(C_{\boldsymbol{M}})$ of (number-valued) smooth
tensor fields $C_{\boldsymbol{M}} = C_{\boldsymbol{M}ab}$ on $\boldsymbol{M}$,
and write $C_{\boldsymbol{M}}(f) = \int_{\boldsymbol{M}} C_{\boldsymbol{M}ab} f^{ab}
\,d{\rm vol}_{\boldsymbol{M}}$ where $d{\rm vol}_{\boldsymbol{M}}$ denotes the
metric-induced volume form on $\boldsymbol{M}$. If $(C_{\boldsymbol{M}})$ is
not only local covariant, but also symmetric and has vanishing divergence, then
we may set
$$ {\sf T}^{[2]}_{\boldsymbol{M}}(f) =  {\sf T}^{[1]}_{\boldsymbol{M}}(f) + C_{\boldsymbol{M}}(f)
{\boldsymbol{1}} $$
(where $\boldsymbol{1}$ is the unit of $\mathscr{A}(\boldsymbol{M})$)
to obtain in this way another choice, $({\sf T}^{[2]}_{\boldsymbol{M}})$, of a
stress energy tensor for $\boldsymbol{\Phi}$ which is as good as the previous one
since it satisfies the conditions {\bf LCSE-1...4} equally well.
Again under quite general conditions, we may expect that this is actually the complete
freedom for the stress energy tensor that is left by the above conditions, in particular
the freedom should in fact be given by a multiple of unity (be state-independent). Since
the $C_{\boldsymbol{M}}$ must be local covariant and divergence-free, they should
be local functionals of the spacetime metric of $\boldsymbol{M}$, and one can prove
this upon adding some technical conditions. 

The freedom which is left by the conditions {\bf LCSE-1...4} for the quantized stress
energy tensor of a local covariant quantum field theory can be traced back to the
circumstance that even in such a simple theory as the linear scalar field,
the stress energy tensor is a renormalized quantity. Let us explain this briefly, using the
example of the minimally coupled linear scalar field, following
the path largely developed by Wald \cite{Wald78,WaldQFTCST}. On any spacetime, the classical
minimally coupled linear scalar field $\phi$ obeys the field equation
\begin{equation} \label{waveeqn}
\nabla^a\nabla_a\phi + m^2\phi = 0 \,.
\end{equation}
 The stress energy tensor for a 
classical solution $\phi$ of the field equation can be presented in the form
$$ T_{ab}(x) = \lim_{y \to x} P_{ab}(x,y;\nabla_{(x)},\nabla_{(y)})\phi(x)\phi(y) $$
(where $x,y$ are points in spacetime) with some partial differential operator 
$P_{ab}(x,y;\nabla_{(x)},\nabla_{(y)})$. This serves as starting point for defining
${\sf T}_{\boldsymbol{M}ab}$ through replacing $\phi$ by the quantized linear
scalar field $\Phi_{\boldsymbol{M}}$. However, one finds already in Minkowski spacetime
that upon performing this replacement, the behaviour of the resulting expression
is singular in the coincidence limit $y \to x$. As usual with non-linear
expressions in a quantum field at coinciding points, one must prescribe
a renormalization procedure in order to obtain a well-defined quantity. In Minkowski
spacetime, this is usually achieved by normal ordering with respect to the vacuum.
This makes explicit reference to the vacuum state in Minkowski spacetime, the
counterpart of which is not available in case of generic spacetimes. Therefore,
one proceeds in a different manner on generic curved spacetimes (but in Minkowski
spacetime, the result coincides with what one obtains from normal ordering, up
to a renormalization freedom). 

Since one is mainly interested in expectation values
of the quantized stress energy tensor ---
as these are the quantities entering on the right hand side of the
semiclassical Einstein equations, discussed in the next section ---
 one may concentrate on a class of states 
$\mathcal{S}(\boldsymbol{M})$ on $\mathscr{A}(\boldsymbol{M})$ for which the
expectation value of the stress energy tensor can be defined as unambiguously 
as possible, and in a manner consistent with the conditions {\bf LCSE-1...4} above.
Defining only the expectation value of the stress energy tensor has the additional
benefit that one need not consider the issue if, or in which sense, 
${\sf T}_{\boldsymbol{M}}(f)$ is contained in $\mathscr{A}(\boldsymbol{M})$ (or
its suitable extensions). Of course, it has the drawback that non-linear 
expressions in  ${\sf T}_{\boldsymbol{M}}(f)$, as they would appear in the
variance of the stress energy, need extra definition, but we may regard this
as an additional issue.  Following the idea above for defining the quantized
stress energy tensor, one can see that the starting point for the expectation
value of the stress energy tensor in a state $\omega_{\boldsymbol{M}}$ on
$\mathscr{A}(\boldsymbol{M})$ is the corresponding two-point function
$$ w_{\boldsymbol{M}}(x,y) = \omega_{\boldsymbol{M}}(\Phi_{\boldsymbol{M}}(x)
\Phi_{\boldsymbol{M}}(y)) \,,$$
written here symbolically as a function, although properly it is a distribution on
$C_0^\infty(\boldsymbol{M} \times \boldsymbol{M})$. One now considers a 
particular set $\mathcal{S}_{\mu sc}(\boldsymbol{M})$, the states on 
$\mathscr{A}(\boldsymbol{M})$ whose two-point functions fulfill the 
{\it microlocal spectrum condition} \cite{Rad,BFK,SVW}. The microlocal spectrum condition
specifies the wavefront set of the distribution $w_{\boldsymbol{M}}$ in
a particular, asymmetric way, which is reminiscent of the spectrum condition
for the vacuum state in Minkowski spacetime. Using the transformation properties
of the wavefront set under diffeomorphisms, one can show (1) 
the sets $\mathcal{S}_{\mu sc}(\boldsymbol{M})$ transform contravariantly under
arrows $\boldsymbol{M} \overset{\psi}{\longrightarrow} \boldsymbol{N}$ in 
{\sf Loc}, meaning that every state in $\mathcal{S}_{\mu sc}(\boldsymbol{N})$
restricts to a state of $\mathcal{S}_{\mu sc}(\psi(\boldsymbol{M}))$ and
$\mathcal{S}_{\mu sc}(\psi(\boldsymbol{M})) \circ \mathscr{A}(\psi) =
\mathcal{S}_{\mu sc}(\boldsymbol{M})$, and moreover (2)   
that the microlocal
spectrum condition is equivalent to the {\it Hadamard condition} on $w_{\boldsymbol{M}}$
\cite{Rad,San}. This condition says \cite{KayWald} that the two-point function
 $w_{\boldsymbol{M}}$ splits in the form
$$ w_{\boldsymbol{M}}(x,y) = \mbox{\sc h}_{\boldsymbol{M}}(x,y)  + u(x,y) $$
where $\mbox{\sc h}_{\boldsymbol{M}}$ is a Hadamard parametrix
for the wave-equation \eqref{waveeqn}, and
$u$ is a $C^\infty$ integral kernel. Consequently, 
the singular behaviour of $\mbox{\sc h}_{\boldsymbol{M}}$ and hence of
$w_{\boldsymbol{M}}$ is determined by the spacetime geometry of $\boldsymbol{M}$
and is state-independent within the set $\mathcal{S}_{\mu sc}(\boldsymbol{M})$
whereas different states are distinguished by different smooth terms $u(x,y)$.
With respect to such a splitting of a two-point function,
one can then define the expectation value of the stress energy tensor in 
a state $\omega_{\boldsymbol{M}}$ in $\mathcal{S}(\boldsymbol{M})$ as
\begin{align} \label{Ttilde}
\langle \tilde{\sf T}_{\boldsymbol{M}ab}(x) \rangle_{\omega_{\boldsymbol{M}}}
 & = \lim_{y \to x} P_{ab}(x,y;\nabla_{(x)},\nabla_{(y)}) \left( w_{\boldsymbol{M}} -
 \mbox{\sc h}_{\boldsymbol{M}} \right) \,.
\end{align} 
Note that $\langle \tilde{\sf T}_{\boldsymbol{M}ab}(x) \rangle_{\omega_{\boldsymbol{M}}}$
is, in fact, smooth in $x$ and therefore is a $C^\infty$ tensor field on $\boldsymbol{M}$.
The reason for the appearance of the twiddle on top of the symbol for the
just defined stress energy tensor is due to the circumstance that with this definition,
te stress energy tensor is (apart from exceptional cases) not divergence-free. However,
Wald \cite{Wald78,WaldQFTCST} has shown that this can be repaired by subtracting the
divergence-causing term. More precisely, there is a smooth function $Q_{\boldsymbol{M}}$
on $\boldsymbol{M}$,
constructed locally from the metric of $\boldsymbol{M}$ (so that the family
$(Q_{\boldsymbol{M}})$ is local covariant), such that
\begin{align} \label{TminusQ} 
\langle {\sf T}_{\boldsymbol{M}ab}(x) \rangle_{\omega_{\boldsymbol{M}}}  & =
 \langle \tilde{\sf T}_{\boldsymbol{M}ab}(x) \rangle_{\omega_{\boldsymbol{M}}}
 - Q_{\boldsymbol{M}}(x) g_{ab}(x)\,,
\end{align} 
 where $g_{ab}$ denotes the spacetime metric of $\boldsymbol{M}$,
defines an expectation value of the stress energy tensor which is divergence-free.
Moreover, with this definition, the conditions {\bf LCSE-1...4} hold when
interpreted as valid for expectation values of states fulfilling the Hadamard condition
\cite{WaldQFTCST,BFV} (possibly after suitable symmetrization in order to obtain
{\bf LCSE-4}). This may actually be seen as one of the main motivations for
introducing the Hadamard condition in \cite{Wald78}, and in the same paper, Wald
proposed conditions on the expectation values of the stress energy tensor which
are variants of our {\bf LCSE-1...4} above. Wald \cite{Wald78,WaldQFTCST}
also proved that, if there are two
differing definitions for the expectation values of the stress energy tensor
complying with the conditions, then the
difference is given by a state-independent, local covariant family $(C_{\boldsymbol{Ma}b})$
of smooth, symmetric, divergence-free tensor fields, in complete analogy to our
discussion above. Actually, the formulation of the local covariance condition for
the expectation values of the stress energy tensor in \cite{WaldQFTCST} was an important
starting point for the later development of local covariant quantum field theory, so
the agreement between Wald's result on the freedom in defining 
the stress energy tensor and ours, discussed above, are in no way coincidential.  

How does this freedom come about? To see this, note that only the singularities of
the Hadamard parametrix
$\mbox{\sc h}_{\boldsymbol{M}}$ are completely fixed. One has the freedom of 
altering the smooth contributions to that Hadamard parametrix, and as long as
this leads to an expected stress energy tensor which is still consistent with conditions
{\bf LCSE-1...4}, this yields an equally good definition of 
$\langle {\sf T}_{\boldsymbol{M}ab}(x) \rangle_{\omega_{\boldsymbol{M}}}$.
Even requiring as in \cite{Wald78,WaldQFTCST} that the expectation value of
the stress energy tensor agrees in Minkowski spacetime with the expression obtained
by normal ordering --- implying that in Minkowski spacetime, the stress energy
expectation value of the vacuum vanishes --- doesn't fully solve the problem. For
if another definition is chosen so that the resulting difference term $C_{\boldsymbol{M}ab}$ is made
of curvature terms,  that difference vanishes on Minkowski spacetime. The problem
could possibly be solved if there was a distinguished state $\boldsymbol{\omega} =
(\omega_{\boldsymbol{M}})$ for which the expectation value of the stress energy tensor
could be specified in some way, but as was mentioned before, the most likely candidate
for such a state, the invariant state, doesn't exist (at least not as a state fulfilling the
microlocal spectrum condition). This means that apparently the setting of local
covariant quantum field theory does not, at least without using further ingrediences,
specify intrinsically the absolute value of the local stress energy content of a 
quantum field state, since the ambiguity of being able to add a difference term 
$C_{\boldsymbol{M}ab}$ always remains. In fact, this difference term ambiguity has the
character of a renormalization ambiguity since it occurs in the process of renormalization
(by means of discarding the singularities of a Hadamard parametrix), as is typical in
quantum field theory. The fact that local covariant quantum field theory  does not
fix the local stress energy content may come as a surprise --- and disappointment --- since
it differs from what one would be inclined to expect from quantum field theory on
Minkowski spacetime with a distinguished vacuum state. This just goes to show that
one cannot too naively transfer concepts from quantum field theory on Minkowski
spacetime to local covariant quantum field theory (a fact which, incidentally, has 
already been demonstrated by the Hawking- and Unruh-effects). 

Another complication needs to be addressed: Anomalies of the quantized stress energy tensor.
It is not only that the value of the (expectation value of) the quantized stress energy tensor
isn't fixed on curved spacetime due to the appearance of metric- or curvature-dependent
renormalization ambiguities, but there are also curvature-induced anomalies. Recall that
in quantum field theory one generally says that a certain quantity is subject to an
anomaly if the quantized/renormalized quantity fails to feature a property ---
mostly, a symmetry property --- which is fulfilled for the counterpart of that quantity in
classical field theory. In the case of the stress energy tensor, this is the trace anomaly
or conformal anomaly. Suppose that $\phi$ is a $C^\infty$ solution of the conformally
coupled, massless linear scalar wave equation, i.e.\ 
$$ (\nabla^a \nabla_a + \frac{1}{6} R)\phi = 0$$
on a globally hyperbolic spacetime $\boldsymbol{M}$ with scalar curvature $R$. Then
the classical stress energy tensor of $\phi$ has vanishing trace, $T^{a}{}_{a} = 0$.
This expresses the conformal invariance of the equation of motion. However, as was shown
by Wald \cite{WaldTrAn}, it is not possible to have the vanishing trace property for the
expectation value of the stress energy tensor in the presence of curvature if one
insists on the stress energy tensor having vanishing divergence: Under these
circumstances, one necessarily finds
$$ \langle {\sf T}_{\boldsymbol{M}}^a{}_a \rangle_{\omega_{\boldsymbol{M}}} = -4 Q_{\boldsymbol{M }}\,.$$
Note that this is independent of the choice of Hadamard state $\omega_{\boldsymbol{M}}$.
The origin of this trace-anomaly lies in having subtracted the divergence-causing
term $Q g_{ab}$ in the definition of the renormalized stress energy tensor. In a sense,
non-vanishing divergence of the renormalized stress energy tensor could also be viewed as an
anomaly, so in the case of the conformally coupled, massless quantized linear scalar,
one can trade the non-vanishing ``divergence anomaly'' of the renormalized stress energy tensor
for the trace anomaly. The condition {\bf LCSE-4} assigns higher priority to vanishing divergence,
whence one has to put up with the trace anomaly.

One may wonder why the features of the quantized/renormalized stress energy tensor
have been discussed here at such an extent while our main topic are local thermal
thermal equilibrium states. The reason is that the ambiguities and anomalies 
by which the definition of the stress energy tensor for quantum fields in curved
spacetime is plagued will also show up when trying to generalize the concept of 
local thermal equilibrium in quantum field theory in curved spacetime. Furthermore,
the ambiguities and anomalies of the stress energy tensor do play a role in semiclassical
gravity (and semiclassical cosmology), and we will come back to this point a bit later.
\section{Local Thermal Equilibrium}
\subsection{LTE States on Minkowski Spacetime}
Following the idea of Buchholz, Ojima and Roos \cite{BOR},
local thermal equilibrium (LTE, for short) states are states of a quantum
field for which local, intensive thermal quantities, such as --- most prominently ---
temperature and pressure can be defined, and take the
values they would assume for a thermal equlibribum state. Here, ``local'' means, in fact, at
a collection of spacetime points. We will soon be more specific about this.

Although the approach of \cite{BOR} covers also interacting fields, for the purposes
of this contribution we will restrict attention to the quantized linear scalar field;
furthermore, we start by introducing the concept of LTE states on Minkowski
spacetime. Consider the quantized massive linear field $\Phi_0$ on Minkowski
spacetime, in its usual vacuum representation, subject to the field equation
$(\Box + m^2)\Phi_0 = 0$ (to be understood in the sense of operator-valued
distributions). Now, at each point $x$ in Minkowski spacetime, one would
like to define a set of  ``thermal observables'' $\boldsymbol{\Theta}_x$ formed
by observables which are sensitive to intensive thermal quantities at $x$.
If that quantity is, e.g., temperature, then the corresponding quantity in
$\boldsymbol{\Theta}_x$ would be a thermometer ``located'' at spacetime
point $x$. One may wonder if it makes physical sense to idealize a thermometer
as being of pointlike ``extension'' in space and time, but as discussed in \cite{BOR},
this isn't really a problem. 

What are typical elements of $\boldsymbol{\Theta}_x$? Let us look at a very particular
example. Assume that $m= 0$, so we have the massless field. Moreover, fix some Lorentzian frame
consisting of a tetrad $(e^a_0,e^a_1,e^a_2,e^a_3)$ of Minkowski vectors such that
$\eta_{ab}e^a_\mu e^b_\nu  = \eta_{\mu\nu}$ and with $e_0^a$ future-directed. Relative to these
data, let $\omega_{\beta,e_0}$ denote the global thermal equilibrium state ---
i.e.\ KMS-state --- with respect to the time-direction $e_0 \equiv e_0^a$ at
inverse temperature $\beta > 0$ \cite{Haag}. Evaluating
$:\Phi_0^2:(x)$, the Wick-square of $\Phi_0$ at the spacetime point $x$, in the KMS state gives
$$ \langle : \Phi_0^2:(x) \rangle_{\omega_{\beta,e_0}} = \frac{1}{12 \beta^2}\,. $$
Recall that the Wick square is defined as 
\begin{equation} \label{Wick1}
   :\Phi_0^2:(x) = \lim_{\zeta \to 0} \Phi_0(q_x(\zeta))\Phi_0(q_x(-\zeta)) -
                                     \langle \Phi_0(q_x(\zeta))\Phi_0(q_x(-\zeta)) \rangle_{\rm vac}
\end{equation}
where $\langle \,.\,\rangle_{\rm vac}$ is the vacuum state, and we have set
\begin{equation} \label{whatisq}
 q_x(\zeta) = x + \zeta
\end{equation}
for Minkowski space coordinate vectors $x$ and $\zeta$, 
tacitly assuming that $\zeta$ isn't zero or lightlike. 
Thus, $:\Phi_0^2:(x)$ is an observable localized at $x$; strictly speaking,
without smearing with test functions with respect to $x$, it 
isn't an operator, but a quadratic form. Evaluating $:\Phi_0^2:(x)$ in the global 
thermal equilibrium state yields a monotonous function of the temperature.
This is also the case for mass $m > 0$, only the monotonous function is a
bit more complicated. So $:\Phi_0^2:(x)$ can be taken as a ``thermometer observable''
at $x$. As in our simple model the KMS-state for $\Phi_0$ is homogeneous
and isotropic, $\langle : \Phi_0^2:(x) \rangle_{\omega_{\beta,e_0}}$ is, in fact,
independent of $x$. Let us abbreviate the Wick-square ``thermometer observable'' by
\begin{equation} \label{Wicki}
 \boldsymbol{\vartheta}(x) = :\Phi_0^2:(x) \,.
\end{equation} 
Starting from the Wick-square one can, following \cite{BOR}, form many more elements of
$\boldsymbol{\Theta}_x$. The guideline was that these elements should be be sensitive to
intensive thermal quantitites at $x$. This can be achieved by forming the {\it balanced derivatives
of the Wick-square of order $n$}, defined as 
\begin{align}                                  
\eth_{\mu_1\ldots\mu_n}& :\Phi_0^2:(x) \\ = & \lim_{\zeta \to 0} 
        \partial_{\zeta^{\mu_1}} \cdots \partial_{\zeta^{\mu_n}} \Big{(}
                         \Phi_0(q_x(\zeta))\Phi_0(q_x(-\zeta))  
                                         - 
                                     \langle \Phi_0(q_x(\zeta))\Phi_0(q_x(-\zeta)) \rangle_{\rm vac} 
 \Big{)} \nonumber
\end{align}
Prominent among these is the second balanced derivative because 
\begin{equation} \label{theen}
\boldsymbol{\varepsilon}_{\mu\nu}(x) = - \frac{1}{4} \eth_{\mu\nu} :\Phi_0^2:(x)  
\end{equation}
is the {\it thermal stress energy tensor} which in the KMS state $\omega_{\beta,e_0}$ 
takes on the values\footnote{We caution the reader that in previous publications on LTE states,
$\vartheta$ and $\varepsilon_{\mu\nu}$ are always used to denote the expectation values of our
$\boldsymbol{\vartheta}$ and $\boldsymbol{\varepsilon}$ in LTE states. We hope that our use of
bold print for the thermal observables is sufficient to distinguish the thermal observables
from their expectation values in LTE states.}
\begin{align}
\varepsilon_{\mu \nu}(x) & =  
\langle \boldsymbol{\varepsilon}_{\mu\nu}(x)\rangle_{\omega_{\beta e_0}}  = \frac{1}{(2\pi )^3} \int_{\mathbb{R}^3} 
 \frac{p_\mu p_\nu}{({\rm e}^{\beta p_0} -1)p_0}\, d^3p
\end{align}
where $p_\mu = p_a e^a_\mu$ are the covariant coordinates of $p$ with respect to the chosen
Lorentz frame, and $p_0 = |\underline{p}|$ with $(p_\mu) = (p_0,\underline{p})$.
Again, this quantity is independent of $x$ in the unique KMS state of our quantum field model.

 Notice that the values of $\varepsilon_{\mu\nu}$ depend not
only on the inverse temperature, but also on the time-direction $e_0$ of the Lorentz-frame with respect to
which $\omega_{\beta,e_0}$ is a KMS-state. Since the dependence of the thermal quantities on 
$\beta$ and $e_0$ in expectation values of $\omega_{\beta,e_0}$ is always a function of 
the timelike, future-directed vector $\boldsymbol{\beta} = \beta \cdot e_0$,
it is hence useful to label the KMS states correspondingly as $\omega_{\boldsymbol{\beta}}$, and to define
LTE states with reference to $\boldsymbol{\beta}$.
\begin{defn} 
Let $\omega$ be a state of the linear scalar field $\Phi_0$ on Minkowski spacetime,
and let $N \in \mathbb{N}$.
\begin{itemize}
\item[(i)]
Let $D$ be a subset of Minkowski spacetime and let 
$\boldsymbol{\beta}: x \mapsto \boldsymbol{\beta}(x)$
be a (smooth, if $D$ is open) map assigning to each $x \in D$ a future-directed timelike vector 
$\boldsymbol{\beta}(x)$. Then we say that {\sl $\omega$ is a local thermal equilibrium state of order $N$
at sharp temperature for the temperature vector field $\boldsymbol{\beta}$} (for short, $\omega$ is
a $[D,\boldsymbol{\beta},N]$-LTE state) if 
\begin{align} \label{lte-s}
 \langle \eth_{\mu_1 \cdots \mu_n} :\Phi_0^2:(x) \rangle_\omega & =
  \langle \eth_{\mu_1 \cdots \mu_n} :\Phi_0^2:(0) \rangle_{\boldsymbol{\beta}(x)}
\end{align}
holds for all $x \in D$  and $0 \le n \le N$. Here, we have written $\langle \,. \rangle_{\boldsymbol{\beta}(x)} =
\omega_{\boldsymbol{\beta}(x)}(\,.\,)$ for the KMS-state of $\Phi_0$ defined with respect to 
the timelike, future-directed vector $\boldsymbol{\beta}(x)$. The balanced derivatives of
the Wick square on the right hand side are evaluated at the spacetime point $0$; since
the KMS state on the right hand side is homogeneous and isotropic, one has
the freedom to make this choice.  
\item[(ii)] Let $D$ be a subset of Minkowski spacetime and let $m: x \mapsto \varrho_x$ be a map
which assigns to each $x \in D$ a probability measure compactly supported on $V_+$, the open set
of all future-directed timelike Minkowski vectors. It will be assumed that the map is smooth
if $D$ is open. Then we say that {\sl $\omega$ is a local thermal equilibrium state of order
$N$ with mixed temperature distribution $\varrho$} (for short, $\omega$ is a 
$[D,\varrho,N]$-LTE state) if 
\begin{align} \label{lte-m}
 \langle \eth_{\mu_1 \cdots \mu_n} :\Phi_0^2:(x) \rangle_\omega & =
  \langle \eth_{\mu_1 \cdots \mu_n} :\Phi_0^2:(0) \rangle_{\varrho_x}
\end{align}
holds for all $x \in D$ and $0 \le n \le N$, where
\begin{align} \label{average}
\langle \eth_{\mu_1 \cdots \mu_n} :\Phi_0^2:(0) \rangle_{\varrho_x} & =
\int_{V_+} \langle \eth_{\mu_1 \cdots \mu_n} :\Phi_0^2:(0) \rangle_{\boldsymbol{\beta}'} \,d{\varrho_x}
(\boldsymbol{\beta}')
\end{align}
\end{itemize}
\end{defn}  
In this definition, $\Phi_0$ can more generally also be taken as the linear scalar 
field with a finite mass different from zero. 
Let us discuss a couple of features of this definition and some results related to it.
\\[6pt]
{\bf (A)} \quad The set of local thermal observables used for testing thermal properties of $\omega$
in (\ref{lte-s}) and (\ref{lte-m})
is $\boldsymbol{\Theta}_x^{(N)}$, formed by the Wick-square and all of its balanced derivatives
up to order $N$. One can generalize the condition to unlimited order of balanced derivatives,
using as thermal observables the set $\boldsymbol{\Theta}_x^{(\infty)} = \bigcup_{N = 1}^{\infty}
\boldsymbol{\Theta}_x^{(N)}$. 
\\[6pt]
{\bf (B)} \quad The condition (\ref{lte-s}) demands that at each $x$ in $D$, the expectation values
of thermal observables in $\boldsymbol{\Theta}_x^{(N)}$ evaluated in $\omega$ and in 
$\langle\,.\,\rangle_{\boldsymbol{\beta}(x)}$ coincide; in other words, with respect to these
thermal observables at $x$, $\omega$ looks just like the thermal equilibrium state
$\langle\,.\,\rangle_{\boldsymbol{\beta}(x)}$. Note that $\boldsymbol{\beta}(x)$ can vary with
$x$, so an LTE state can have a different temperature and a different ``equilibrium rest frame'',
given by the direction of $\boldsymbol{\beta}(x)$, at each $x$. 
\\[6pt]
{\bf (C)} \quad
With increasing order $N$, the sets $\boldsymbol{\Theta}^{(N)}_x$ become larger; so the higher
the order $N$ for which the LTE condition is fulfilled, the more can the state $\omega$ be regarded
as coinciding with a thermal state at $x$. In this way, the maximum order $N$ for which 
(\ref{lte-s}) is fulfilled provides a measure of the
deviation from local thermal equilibrium (and similarly, for the condition (\ref{lte-m})).  
\\[6pt]
{\bf (D)} \quad  Condition (\ref{lte-s}) demands that, on $\boldsymbol{\Theta}_x^{(N)}$, $\omega$
coincides with a thermal equilibrium state at sharp temperature, and sharp thermal rest-frame.
Condition (\ref{lte-m}) is less restrictive, demanding only that $\omega$ coincides with
a mixture of thermal equilibrium states, described by the probability measure $\varrho_x$, on
the thermal observables $\boldsymbol{\Theta}_x$. Of course, (\ref{lte-s}) is a special case
of (\ref{lte-m}).
\\[6pt] 
{\bf (E)} \quad Clearly, each global thermal equilibrium state, or KMS-state, $\omega_{\boldsymbol{\beta}}$
is an LTE state, with constant inverse temperature vector $\boldsymbol{\beta}$. The interesting
feature of the definition of LTE states is that $\boldsymbol{\beta}(x)$ can vary with $x$ in
spacetime, and the question of existence of LTE states which are not global KMS states arises.
In particular, fixing the order $N$ and the spacetime region $D$, which functions $\boldsymbol{\beta}(x)$
or $\varrho_x$ can possibly occur? They surely cannot be completely arbitrary, particularly for open
$D$, since the Wick-square of $\Phi_0$ and its balanced derivatives are subject to 
dynamical constraints which are a consequence of the equation of motion for $\Phi_0$. 
We put on record some of the results which have been established so far.
\\[10pt]
{\bf The hot bang state} \cite{BOR,BuHB}
\\
 A state of $\Phi_0$ (zero mass case) 
which is an LTE state with a 
variable, sharp inverse temperature vector field on the open forward lightcone $V_+$
was constructed in \cite{BOR} and further investigated in \cite{BuHB}.
This state is called the hot bang state $\omega_{\rm HB}$, and for $x,y$ in $V_+$,
its two-point function $w_{\rm HB}$ has the form
\begin{align} \label{2pHB}
w_{\rm HB}(x,y) &= \frac{1}{(2 \pi)^{3}} \int_{\mathbb{R}^4}
     {\rm e}^{- i (x - y)^\mu p_\mu} \epsilon(p_0) \delta(p^\mu p_\mu) 
     \frac{1}{1 - {\rm e}^{- a ((x - y)^\mu p_\mu)}} d^4p
\end{align}
where $a > 0$ is some parameter, and $\epsilon$ denotes the
sign function. To compare, the two-point function of a KMS-state
with constant inverse temperature vector $\boldsymbol{\beta}'$ is given by
\begin{align} \label{2pKMS}
w_{\boldsymbol{\beta}'}(x,y) &= \frac{1}{(2 \pi)^{3}} \int_{\mathbb{R}^4}
     {\rm e}^{- i (x - y)^\mu p_\mu} \epsilon(p_0) \delta(p^\mu p_\mu) 
     \frac{1}{1 - {\rm e}^{- \boldsymbol{\beta}'{}^\mu p_\mu}} d^4p\,.
\end{align}
Upon comparison, one can see that the inverse temperature vector field
of $\omega_{\rm HB}$ has an inverse temperature vector field
$$ \boldsymbol{\beta}(x) = 2 a x\,, \quad x \in V_+\,.$$
Thus, the temperature diverges at the boundary of $V_+$, with the thermal rest
frame tilting lightlike; moreover, the temperature decreases away from the 
boundary with increasing coordinate time $x^0$. This behaviour is sketched in 
Figure 1.
\\[4pt] 
\begin{center}
\includegraphics[totalheight=3.0cm]{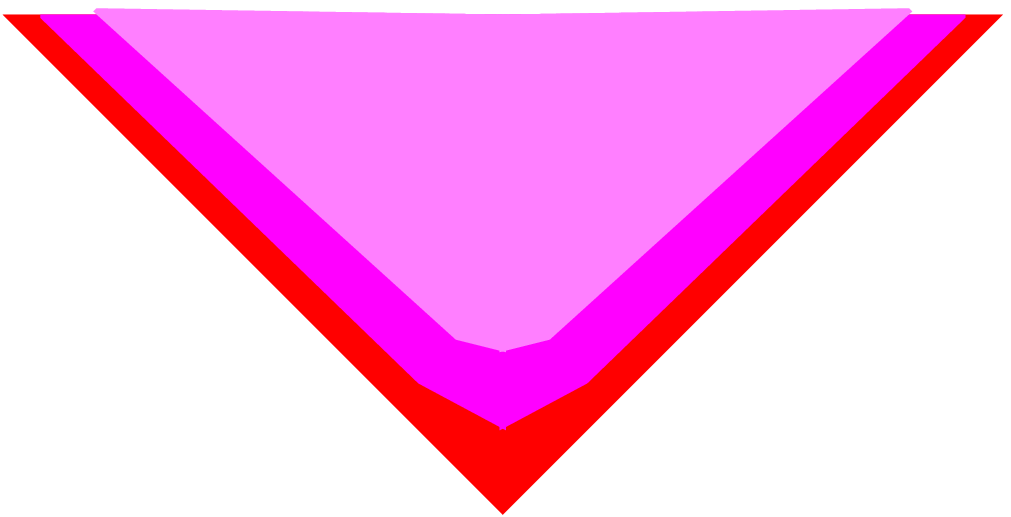}
\\[4pt]
{\bf Figure 1: Sketch of temperature distribution of the hot bang state.}
\end{center}
${}$\\
It also worth mentioning that the hot bang state provides an example of a
state where the thermal stress-energy tensor
$\varepsilon_{\mu\nu}(x) = -(1/4) \langle \eth_{\mu \nu} : \Phi_0^2:(x) \rangle_{\rm HB}$
deviates from the expectation values of the full stress energy tensor 
$\langle {\sf T}_{\mu\nu}(x) \rangle_{\rm HB}$.
In fact, as discussed in \cite{BOR}, for the stress energy tensor of the massless linear scalar field in 
Minkowski spacetime one finds generally 
$$ {\sf T}_{\mu\nu}(x) = \boldsymbol{\varepsilon}_{\mu\nu} + \frac{1}{12} (\partial_\mu \partial_\nu
 - \eta_{\mu\nu} \Box) : \Phi_0^2:(x)\,, $$
and thus one obtains, for the hot bang state,
$$ \varepsilon^{({\rm HB})}_{\mu\nu}(x) =
\langle \boldsymbol{\varepsilon}_{\mu\nu}(x) \rangle_{\rm HB} =
\frac{\pi^2}{1440a^4} \frac{4 x_\mu x_\nu - \eta_{\mu\nu} x^\lambda x_\lambda}{(x^\kappa x_\kappa)^3}\,,$$
whereas
$$ \langle {\sf T}_{\mu\nu}(x) \rangle_{\rm HB} = {\varepsilon}^{({\rm HB})}_{\mu\nu}(x)
 + \frac{1440a^2}{288 \pi^2} {\varepsilon}^{({\rm HB})}_{\mu\nu}(x) \,.$$
The first term is due to the thermal stress energy, while the second term is a convection
term which will dominate over the thermal stress energy when the parameter $a$ exceeds 
1 by order of magnitude. Thus, for non-stationary LTE states, the expectation value of the
full stress energy tensor can in general not be expected to coincide with the thermal
stress energy contribution due to transport terms which are not seen in $\varepsilon_{\mu\nu}$. 
\\[10pt]
{\bf Maximal spacetime domains for nontrivial LTE states} \cite{BuHB} 
\\
As mentioned before, the inverse temperature vector field for an LTE state cannot
be arbitrary since the dynamical behaviour of the quantum field imposes constraints.
A related, but in some sense stronger constraint
which appears not so immediate comes in form of the following result which was 
established by Buchholz in \cite{BuHB}: Suppose that $\omega$ is a $[D,\boldsymbol{\beta},\infty]$-LTE
state, i.e.\ an LTE state of infinite order with sharp inverse temperature vector field
$\boldsymbol{\beta}$. If $D$, the region on which $\omega$ has the said LTE-property,
contains a translate of $V_+$, then for $\boldsymbol{\beta}$ to be non-constant it
is necessary that $D$ is contained in some timelike simplicial cone, i.e.\ an
intersection of characteristic half-spaces (which means, in particular, that 
$D$ cannot contain any complete timelike line).  
\\[10pt]
{\bf Existence of non-trivial mixed temperature LTE states} \cite{Solv}
\\
The hot bang state mentioned above has been constructed for the
case of the massless linear scalar field. It turned out to be more difficult
to construct non-trivial LTE states for the massive linear scalar field, and the 
examples known to date for the massive case are mixed temperature LTE
states. Recently, Solveen \cite{Solv}  proved a general result on the existence
of non-trivial mixed temperature LTE states: Given any compact subset $D$
of Minkowski spacetime, there are 
 non-constant probability measure-valued functions
$x \mapsto \varrho_x$, $x \in D$, together with states $\omega$ which are 
$[D,\varrho,\infty]$-LTE states. 
\newpage \noindent
{\bf LTE condition as generalization of the KMS condition} \cite{NPRV}
\\
The choice of balanced derivatives of the Wick-square as thermal observables
fixing the LTE property may appear, despite the motivation given in \cite{BOR}, as
being somewhat arbitrary, so that one would invite other arguments for their
prominent role in setting up the LTE condition. One attempt in this direction
has been made by Schlemmer who pointed at a relation between an Unruh-like
detector model and balanced derivatives of the Wick-square \cite{SchlDetc}.
On the other hand, recent work by Pinamonti and Verch shows that
the LTE condition can be viewed as a generalization of the KMS condition.
The underlying idea will be briefly sketched here, for full details see the
forthcoming publication \cite{NPRV}.
For a KMS-state $\langle \,.\,\rangle_{\beta,e_0}$, let
$$\varphi(\tau) = \varphi_x(\tau) =
 \langle \Phi_0(q(-\frac{1}{2} \tau e_0))\Phi_0( q( +  \frac{1}{2} \tau e_0))\rangle_{\beta,e_0}\,.$$
Then the KMS condition implies that there is a function
$$ f=f_x : S_\beta = \{\tau + i \sigma: \tau \in \mathbb{R},\ 0 < \sigma < \beta\} \to \mathbb{C}$$
which is analytic on the open strip, defined and continuous on the closed strip except at the boundary points
with $\tau = 0$,
such that
$$ \lim_{\sigma\to 0}(\varphi(\tau) - f(\tau + i\sigma)) = 0 \ \ \text{and} \ \
\lim_{\sigma' \to \beta} (\varphi(-\tau) - f(\tau + i \sigma')) = 0 \,. $$
Now $\langle\,.\,\rangle_\omega$ be a (sufficiently regular)
 state for the quantum field $\Phi_0$, and let 
$$\psi(\tau) = \psi_x(\tau) = 
\langle \Phi_0(q_x(-\frac{1}{2} \tau e_0))\Phi_0( q_x(\frac{1}{2} \tau e_0))\rangle_\omega\,.$$
Setting $\boldsymbol{\beta} = \beta e_0$ as before, and taking $D = \{x\}$, i.e.\ the set containing
just the point $x$, it is not difficult to see that 
$\langle\,.\rangle_\omega$ is an
 $[\{x\},\boldsymbol{\beta},N]$-LTE state iff there is a function
$$ f=f_x : S_\beta = \{\tau + i \sigma: \tau \in \mathbb{R},\ 0 < \sigma < \beta\} \to \mathbb{C}$$
which is analytic on the open strip, defined and continuous on the closed strip except
 at the boundary 
points with $\tau = 0$,
such that
\begin{align*}
\lim_{\tau \to 0} & \partial_{\tau}^n \lim_{\sigma \to 0}(\psi(\tau) - f(\tau + i\sigma)) = 0 \ \ \text{and} \\
\lim_{\tau \to 0} & \partial_{\tau}^n\lim_{\sigma' \to \beta}(\psi(-\tau) - f(\tau + \sigma')) = 0
 \ \ (n \le N) \,.
\end{align*}
In this sense, the LTE condition appears as a generalization of the KMS condition.
\\[10pt]
{\bf Structure of the Set of Thermal Observables}
\\
The linearity of the conditions (\ref{lte-s}) and (\ref{lte-m})
implies that they are also fulfilled for linear combinations of elements in
$\boldsymbol{\Theta}_x^{(N)}$. This means that the LTE property of a state extends
to elements of the vector space spanned by $\boldsymbol{\Theta}_x^{(N)}$. That is of
some importance since one can show that ${\rm span}(\boldsymbol{\Theta}^{(N)}_x)$ is dense in
the set of all thermal observables, its closure containing, e.g., the entropy flux density
\cite{BOR,BuHB}. 

Furthermore, the equation of motion of $\Phi_0$ not only provides contraints on the possible
functions $\boldsymbol{\beta}(x)$ or $\varrho_x$ which can occur for LTE states, but
even determine evolution equations for these --- and other --- thermal observables in LTE states
\cite{BOR,BuHB}. Let us indicate this briefly by way of an example.
 For an LTE state, it holds that the trace of
the thermal stress energy tensor vanishes, $\varepsilon^\nu{}_\nu(x) = 0$, due to
the analogous property for KMS states. On the other hand,
from relations between Wick products and their balanced derivatives one obtains 
the equation
$\boldsymbol{\varepsilon}^{\nu}_{\nu}(x) = \partial^{\nu}\partial_{\nu} : \Phi_0^2:(x)$, and
therefore, for any LTE state which fulfills the LTE condition on an open domain, the
differential equation $\partial^{\nu} \partial_{\nu} \vartheta(x) = 0$ must hold, where
$\vartheta(x)$ denotes the expectation values of $\boldsymbol{\vartheta}(x)$, i.e.\
the Wick square, in the LTE state. This is a differential equation for (a function of) 
$\boldsymbol{\beta}(x)$ or $\varrho_x$.  

\subsection{LTE States in Curved Spacetime}

The concept of LTE states, describing situations which are locally approximately
in thermal equilibrium, appears promising for quantum field theory in curved spacetime where
global thermal equilibrium in general is not at hand. This applies in particular to early
stages in cosmological scenarios where thermodynamical considerations play a central role 
for estimating processes which determine the evolution and structures of the Universe in
later epochs. However, usually in cosmology the thermal stress energy tensor is identified with
the full stress energy tensor, and it is also customary to use an ``instantaneous equilibrium''
description of the (thermal) stress energy tensor. In view of our previous discussion, this can
at best be correct in some approximation. Additional difficulties arise from the ambiguities
and anomalies affecting the quantized stress energy tensor, and we will see that such difficulties 
are also present upon defining LTE states in curved spacetime. Let us see how one may proceed in
trying to generalize the LTE concept to curved spacetime, and what problems are met in that attempt. 

We assume that we are given a local covariant quantum field $\boldsymbol{\Phi} = ( \Phi_{\boldsymbol{M}})$
with a local covariant stress energy tensor; we also assume that $\boldsymbol{\Phi}$ admits
global thermal equilibrium states for time-symmetries on flat spacetime. For concreteness, we will
limit our discussion to the case that our local covariant quantum field scalar field
with curvature coupling $\xi$. The parameter $\xi$ can be any real
number, but the most important cases are $\xi = 0$ (minimal coupling) 
or $\xi = 1/6$ (conformal coupling). So $\Phi_{\boldsymbol{M}}$ obeys the field equation
\begin{align} \label{fildeq}
( \nabla^a\nabla_a + \xi R  + m^2) \Phi_{\boldsymbol{M}} & = 0 \,,
\end{align}
where $\nabla$ is the covariant derivative of $\boldsymbol{M}$ and $R$ the associated scalar curvature,
with mass $m \ge 0$.

The central objects in the definition of LTE states in flat spacetime were (i) at each spacetime point
$x$ a set (or family of sets) $\boldsymbol{\Theta}_x^{(N)}$ containing ``thermal observables''
localized at $x$, and (ii) a set of global thermal equilibrium states (KMS-states) serving
as ``reference states'' determining the thermal equilibrium values of elements in
$\boldsymbol{\Theta}_x$. In a generic curved spacetime one encounters the problem that there aren't
any global thermal equilibrium states (unless the spacetime possesses suitable time-symmetries)
and thus there are in general no candidates for the requisite reference states of (ii). To circumvent
this problem, one can take advantage of having assumed that $\boldsymbol{\Phi}$ is a local
covariant quantum field theory. Thus, let $\boldsymbol{M}$ be a globally hyperbolic spacetime, and
let $\Phi_{\boldsymbol{M}}$ be the quantum field on $\boldsymbol{M}$ given by $\boldsymbol{\Phi}$.
There is also a quantum field $\Phi_0$ on Minkowski spacetime given by $\boldsymbol{\Phi}$. One can
use the exponential map ${\rm exp}_x$ in $\boldsymbol{M}$ at $x$ to identify $\Phi_{\boldsymbol{M}} \circ 
{\rm exp}_x$ and $\Phi_0$, and thereby one can push forward thermal equilibrium states
$\omega_{\boldsymbol{\beta}}$ of $\Phi_0$ to thermal reference states on which to evaluate correlations
of $\Phi_{\boldsymbol{M}} \circ {\rm exp}_x$ infinitesimally close to $x$. This way of defining thermal reference
states at each spacetime point $x$ in $\boldsymbol{M}$ has been proposed by 
Buchholz and Schlemmer \cite{BuSchl}. 

As a next step, one needs a generalization of balanced derivatives of the Wick square of $\Phi_{\boldsymbol{M}}$
as elements of the $\boldsymbol{\Theta}^{(N)}_{\boldsymbol{M},x}$ where we used the label
$\boldsymbol{M}$ to indicate that the thermal observables are defined with respect to the spacetime
$\boldsymbol{M}$. So far, mostly the case $N = 2$ has been considered, which is enough to study the
thermal stress energy tensor. In \cite{SchlVer}, the following definition was adopted: Let 
$\omega_{\boldsymbol{M}}$ be a state of $\Phi_{\boldsymbol{M}}$ fulfilling the microlocal
spectrum condition (in other words, $\omega_{\boldsymbol{M}}$ is a Hadamard state), and let
$w_{\boldsymbol{M}}$ denote the corresponding two-point function. We consider the smooth part
$u(x,y) = w_{\boldsymbol{M}}(x,y) - \mbox{\sc h}_{\boldsymbol{M}}(x,y)$ obtained from the two-point function after
subtracting the singular Hadamard parametrix $\mbox{\sc h}_{\boldsymbol{M}}(x,y)$ as in (\ref{Ttilde}). Then we take
its symmetric part $u_+(x,y) = \frac{1}{2}( u(x,y) + u(y,x))$ and define the expectation value of the
Wick square as the coincidence limit
\begin{align} \label{covWick}
\langle : \Phi_{\boldsymbol{M}}^2:(x) \rangle_{\omega_{\boldsymbol{M}}} & = \lim_{y \to x}\, u_+(x,y)\,.
\end{align}
Furthermore, we define the second balanced derivative of the Wick square of $\Phi_{\boldsymbol{M}}$
in terms of expectation values as
\begin{align} \label{twostar}
\langle \eth_{ab} : \Phi_{\boldsymbol{M}}^2:(x)\rangle_{\omega_{\boldsymbol{M}}} & =
\lim_{x' \to x} \Big{(} \nabla_a \nabla_b - \nabla_a \nabla_{b'} - \nabla_{a'} \nabla_b +
\nabla_{a'} \nabla_{b'} \Big{)} u_+(x,x')
\end{align}
where on the right hand side, unprimed indices indicate covariant derivatives with respect to
$x$, whereas primed indices indicate covariant derivatives with respect to $x'$. 
In \cite{SchlVer} it is shown that (\ref{twostar}) amounts to taking the second derivatives
of $u_+({\rm exp}_x(\zeta),{\rm exp}_x(-\zeta))$ with respect to $\zeta$, and evaluating
at $\zeta = 0$.  Note that upon using the symmetric part $u_+$ of $u$ in defining
the Wick product, its first balanced derivative vanishes, as it would on
Minkowski spacetime for the normal ordering definition. The definitions (\ref{covWick}) and
(\ref{twostar}) are referred to as {\it symmetric Hadamard parametrix substraction (SHP)} prescription,
as in \cite{SchlVer}. Note, however, that for the massive case, $m>0$, the SHP prescription differs
from the normal ordering prescription by $m$-dependent universal constants, a fact which
must be taken into account when defining the LTE condition; see \cite{SchlVer} for details.
 
By definition, $\boldsymbol{\Theta}^{(2)}_{\boldsymbol{M},x}$ is then taken to consist of
multiples of the unit operator, $: \Phi_{\boldsymbol{M}}^2:(x)$ and $\eth_{ab} : \Phi_{\boldsymbol{M}}^2:(x)$,
defined according to the SHP prescription. With these ingredients in place, one can attempt the definition
of LTE states on a curved, globally hyperbolic spacetime $\boldsymbol{M}$.

\begin{defn}
Let $\omega_{\boldsymbol{M}}$ be a state of $\Phi_{\boldsymbol{M}}$ fulfilling the microlocal spectrum
condition, and let $D$ be a subset of the spacetime $\boldsymbol{M}$.
\begin{itemize}
\item[(i)]
We say that $\omega_{\boldsymbol{M}}$ is an LTE state with sharp temperature vector field
(timelike, future-directed, and smooth if $D$ is open)
$\boldsymbol{\beta} : x \mapsto \boldsymbol{\beta}(x)$ $(x \in D)$ of order 2 if
\begin{align} \label{e-a}
\langle : \Phi_{\boldsymbol{M}}^2:(x) \rangle_{\omega_{\boldsymbol{M}}} & = \langle : \Phi_{0}^2:(0) \rangle_{\boldsymbol{\beta}(x)} \quad \text{and} \\
\langle \eth_{\mu \nu} : \Phi_{\boldsymbol{M}}^2:(x)\rangle_{\omega_{\boldsymbol{M}}} & = \langle \eth_{\mu\nu} : \Phi_{0}^2:(0)\rangle_{\boldsymbol{\beta}(x)} 
\end{align}
holds for all $x \in D$. On the right hand side there appear the expectation values of the thermal equilibrium state
$\langle \,.\,\rangle_{\boldsymbol{\beta}(x)}$ of $\Phi_0$ on Minkowski spacetime, where the vector $\boldsymbol{\beta}(x) \in T_x{\boldsymbol{M}}$
on the left hand side is identified with a Minkowski space vector $\boldsymbol{\beta}(x)$ on the right hand side via the
exponential map ${\rm exp}_x$, using that ${\rm exp}_x(0) = x$, and the coordinates refer
to a choice of Lorentz frame at $x$. On the right hand side,
we have now used the definition of Wick-product and its second balanced derivative 
according to the SHP prescription.
\item[(ii)]
Let $\varrho: x \mapsto \varrho_x$ $(x \in D)$ be a map from $D$ to compactly supported probability measures in $V_+$
(assumed to be smooth is $D$ is open). We say that $\omega_{\boldsymbol{M}}$ is an LTE state with mixed temperature
distribution $\varrho$ of order 2 if
\begin{align} \label{e-b}
\langle : \Phi_{\boldsymbol{M}}^2:(x) \rangle_{\omega_{\boldsymbol{M}}} & = \langle : \Phi_{0}^2:(0) \rangle_{\varrho(x)} \quad \text{and} \\
\langle \eth_{\mu \nu} : \Phi_{\boldsymbol{M}}^2:(x)\rangle_{\omega_{\boldsymbol{M}}} & = 
\langle \eth_{\mu\nu} : \Phi_{0}^2:(0)\rangle_{\varrho(x)} 
\end{align}
holds for all $x \in D$. The same conventions as in (i) regarding identification of curved spacetime objects (left hand sides)
and Minkowski space objects (right hand sides) by ${\rm exp}_x$ applies here as well. The definition of the $\varrho_x$-averaged objects
is as in (\ref{average}).
\end{itemize}
\end{defn}
Let us discuss some features of this definition and some first results related to it on
generic spacetimes. We will present results pertaining to LTE states on cosmological
spacetimes in the next section.
\\[10pt]
$\boldsymbol{(\alpha)}$ \quad The definition of thermal observables given here is local
covariant since the Wick square and its covariant derivatives are local covariant quantum
fields \cite{HolWald}. In particular, if $\mathscr{A}$ is the functor describing our
local covaraint quantum field $\boldsymbol{\Phi}$, then for any arrow $\boldsymbol{M} 
\overset{\psi}{\longrightarrow} \boldsymbol{N}$ one has 
$\mathscr{A}(\psi)(\boldsymbol{\Theta}^{(2)}_{\boldsymbol{M},x}) =
\boldsymbol{\Theta}^{(2)}_{\boldsymbol{N},\psi(x)}$ (by a suitable extension
of $\mathscr{A}$, see \cite{HolWald}). Obviously, it is desirable
to define  thermal observables in curved spacetimes in such a way as to be local
covariant. Otherwise, they would depend on some global properties of the particular
spacetimes, in contrast to their interpretation as local intensive quantities.
\\[10pt]
$\boldsymbol{(\beta)}$ \quad As is the case for the quantized stress energy tensor, also the
thermal observables in $\boldsymbol{\Theta}_{\boldsymbol{M},x}^{(2)}$, i.e.\ the Wick square of
$\Phi_{\boldsymbol{M}}$ and its second balanced derivative, are subject to renormalization
ambiguities and anomalies. Supposing that choices of
$: \Phi_{\boldsymbol{M}}^{2\,[1]}:(x)$ and $\eth_{\mu \nu} : \Phi_{\boldsymbol{M}}^{2\,[1]}:(x)$
have been made, where $[1]$ serves as label for the particular choice, one has, in principle, the
freedom of redefining these observables by adding suitable quantites depending only on the
local curvature of spacetime (so as to preserve local covariance), like
\begin{align*}
 : \Phi_{\boldsymbol{M}}^{2\,[2]}:(x) & = : \Phi_{\boldsymbol{M}}^{2\,[1]}:(x) + {\sf y}_{\boldsymbol{M}}^{[2][1]}(x)
 \,, \\
 \eth_{\mu \nu} : \Phi_{\boldsymbol{M}}^{2\,[1]}:(x) & = \eth_{\mu \nu} : \Phi_{\boldsymbol{M}}^{2\,[1]}:(x)
  + Y_{\boldsymbol{M}\mu\nu}^{[2][1]}(x) \,.
\end{align*}
Therefore, on curved spacetime the thermal interpretation of Wick square and its balanced derivatives 
depends on the choice one makes here, and it is worth contemplating if there are preferred choices
which may restrict the apparent arbitrariness affecting the LTE criterion.
\\[10pt]
$\boldsymbol{(\gamma)}$ \quad 
Similarly as observed towards the end of the previous Section,
there are differential equations to be fulfilled in order that the
LTE condition can be consistent. For the case of the linear (minimally coupled,
massless) scalar field, Solveen \cite{Solv}
has noted that the condition of  the thermal stress energy tensor 
having vanishing trace in LTE states leads to a differential equation
of the form
\begin{align*}
\frac{1}{ 4} \nabla^a \nabla_a \vartheta(x) + \xi R(x) \vartheta(x) + U(x) & = 0
\end{align*}
where $\vartheta$ is the expectation value of the Wick square in an LTE state,
$R$ is the scalar curvature of the underlying spacetime $\boldsymbol{M}$, and
$U$ is another function determined by the curvature of $\boldsymbol{M}$.
In view of the previous item $\boldsymbol{(\beta)}$, a redefinition of the
Wick square will alter the function $U(x)$ of the differential equation that
must be obeyed by $\vartheta(x)$. The consequences of that possibility are 
yet to be determined. It is worth mentioning that for the Dirac field,
which we don't treat in these proceedings, an analogous consistency
condition leads to an equation which can only be fullfilled provided that --- in this
case --- the first balanced derivative is defined appropriately, i.e.\ with addition
of a distinct curvature term relative to 
the SHP definition of the Wick square \cite{Knospe}. We admit that so far we do not fully understand
the interplay of the LTE condition and the renormalization ambiguity which is
present in the definition of Wick products and their balanced derivatives in
curved spacetime, but hope to address some aspects of that interplay in
greater detail elsewhere \cite{GrKnVe}. 
\\[10pt] 
$\boldsymbol{(\delta)}$
\quad Finally we mention that one can prove so-called ``quantum energy inequalities'',
i.e.\ lower bounds on weighted integrals of the energy density in LTE states
along timelike curves (see
\cite{FewsterReview} for a review on quantum energy inequalities); 
the lower bounds depend on the maximal temperature
an LTE state attains along the curve. 
This holds for a wide range of curvature couplings $\xi$ and
all mass parameters $m$ \cite{SchlVer}.  That is of interest since, while state-independent
lower bounds on weighted integrals of the energy density have been established
for all Hadamard states of the minimally coupled linear scalar field in generic
spacetimes \cite{Fewster}, such a result fails in this generality for the non-minimally
coupled scalar field \cite{FewOst}. We recommend that the reader takes a look at the
references for further information on this circle of questions and their possible
relevance regarding the occurrence of singularities in solutions to the 
semiclassical Einstein equations.

\section{LTE States on Cosmological Spacetimes}
As mentioned previously, one of the domains where one can apply the concept
of LTE states and also examin its utility is early cosmology. Thus, we review
the steps which have been taken, or are currently being taken, in investigating
LTE states in cosmological scenarios.

The central premise in standard cosmology is that one considers phenomena at
sufficiently large scales such that it is a good approximation to assume that,
at each instant of time, the geometry of space (and, in order to be consistent
with Einstein's equations, the distribution of matter and energy) is isotropic
and homogeneous \cite{Weib}. Making for simplicity the additional assumption
(for which there seems to be good observational motivation) that the geometry of
space is flat at each time, one obtains
\begin{align} \label{frw}
 I \times \mathbb{R}^3 \,, \quad & ds^2 = dt^2 - a(t)^2\Big{(} (dx^1)^2 + (dx^2)^2 + (dx^3)^2 \Big{)}
\end{align}
for the general form of spacetime manifold and metric, respectively, in standard
cosmology. Here, $I$ is an open interval hosting the time-coordinate $t$,
and $a(t)$ is a smooth, strictly positive function called the {\it scale factor}.
With the spacetime geometry of the general form (\ref{frw}), the only freedom
is the time-function $a(t)$, to be determined by Einstein's equation together
with a matter model and initial conditions. 

The time-coordinate in (\ref{frw}) is called {\it cosmological time}. Under suitable
(quite general) conditions on $a(t)$ one can pass to a new time-coordinate, called
{\it conformal time},
\begin{align*}
 \eta & = \eta(t) = \int_{t_0}^t \frac{dt'}{a(t')}\, dt'
\end{align*}
for some choice of $t_0$. Setting
\begin{align*}
\Omega(\eta) = a(t(\eta))\,,
\end{align*}
the metric (\ref{frw}) takes the form
\begin{align} \label{conft}
ds^2 & = \Omega(\eta)^2 \Big{(} d\eta^2 - (dx^1)^2 - (dx^2)^2 - (dx^3)^2 \Big{)}
\end{align}
with respect to the conformal time-coordinate, so it is conformally equivalent
to flat Minkowski spacetime.\footnote{For simplicity, we assume here that the 
range of the conformal time-coordinate $\eta$ is all of $\mathbb{R}$; the
variations in the following arguments for the --- important --- case where this is not so should be
fairly obvious.} 

\subsection{Existence of LTE States at Fixed Cosmological Time} \label{SchlRes}

Now let $\Phi_{\boldsymbol{M}}$ be the linear, quantized scalar
field on the cosmological spacetime (\ref{frw}) for some $a(t)$, or equivalently 
on $\mathbb{R}^4$ with metric (\ref{conft}); we assume arbitrary curvature coupling,
as in (\ref{fildeq}). The first question one would like to answer is if there are 
LTE states of order 2 for $\Phi_{\boldsymbol{M}}$. As one might imagine, this is a
very difficult problem, and it seems that the method employed by Solveen \cite{Solv} to
establish existence of non-trivial LTE states on bounded open regions of Minkowski
spacetime cannot be used to obtain an analogous result in curved spacetime. In view
of the constraints on the temperature evolution it seems a good starting point
to see if there are any 2nd order LTE states at some fixed cosmological time, i.e.\
on a Cauchy-surface --- this would postpone the problem of having to establish
solutions to the evolution equations of the temperature distribution. Moreover,
to fit into the formalism, such states have to fulfill the microlocal spectrum
condition, i.e.\ they have to be Hadamard states. Allowing general $a(t)$, this problem
is much harder than it seems, since if one took an ``instantaneous KMS state'' at
some value of cosmological time --- such a state, defined in terms of the Cauchy data
formulation of the quantized linear scalar field, appears as a natural candidate
for an LTE state at fixed time --- then that state is in general not Hadamard if
$a(t)$ is time-dependent. The highly non-trivial problem was solved by Schlemmer
in his PhD thesis \cite{SchlPhD}. The result he established is as follows. 
\begin{thm}
Let $t_1$ be a value of cosmological time in the interval $I$, and let
$e_0^a = (dt)^a$ be the canonical time-vectorfield of the spacetime (\ref{frw}).
Then there is some $\beta_1 > 0$ (depending on $m$, $\xi$ and the behaviour of
$a(t)$ near $t_1$) such that, for each $\beta < \beta_1$, there is a quasifree
Hadamard state of $\Phi_{\boldsymbol{M}}$ which is a
$[\{t_1\} \times \mathbb{R}^3,\beta e_0^a, 2]$-LTE state (at sharp temperature).
\end{thm} 
In other words, there is a 2nd order LTE state at sharp temperature at fixed
cosmological time provided the LTE temperature is high enough. This state will, in general,
not preserve the sharp temperature 2nd order LTE property when evolving it by the field
equation (\ref{fildeq}) in time away from the $t_1$-Cauchy-surface. In Schlemmer's
thesis, this is illustrated by means of a numerical example. Let us take some
 $[\{t_1\} \times \mathbb{R}^3,\beta e_0^a, 2]$-LTE state $\omega$, and define
$$\theta(x) = \langle \boldsymbol{\vartheta}(x)\rangle_\omega\quad \text{and} \quad
 \epsilon_{ab}(x) = \langle \boldsymbol{\varepsilon}_{ab}(x) \rangle_\omega $$
as the expectation values of Wick-square and thermal stress energy tensor
in that state. One can, at each $x$, look for a 2nd order
 LTE state $\omega_{\boldsymbol{\beta}(x)}$ at $x$
such that $\omega_{\boldsymbol{\beta}(x)}(\boldsymbol{\vartheta}(x)) = \theta(x)$.
 Likewise, one can look for 2nd order LTE states
$\omega_{\boldsymbol{\beta}_{\mu\nu}(x)}$ at $x$ with  $\boldsymbol{\beta}_{\mu\nu}(x)
= \beta_{\mu\nu}(x)e^a_0$ and 
$\omega_{\boldsymbol{\beta}_{\mu\nu}(x)}(\boldsymbol{\varepsilon}_{\mu\nu}(x)) = \epsilon_{\mu\nu}(x)$
(note that there is no sum on the indices)
with respect to a tetrad basis at $x$ containing $e_0^a$.
If $\omega$ is itself a 2nd order LTE-state at sharp temperature, then all the
numbers $\beta(x)$ and $\beta_{\mu\nu}(x)$ must coincide, or
rather the associated absolute temperatures
${\rm T}(x) = 1/k\beta(x)$ and ${\rm T}_{\mu\nu}(x) = 1/k\beta_{\mu\nu}(x)$. Otherwise, the
mutual deviation of these numbers can be taken as a measure for the
failure of $\omega$ to be a sharp temperature, 2nd order LTE state at $x$. 
Schlemmer has investigated such a case, choosing $\xi = 0.1$ and $m=1.5$
in natural units, and $a(t) = {\rm e}^{Ht}$ with $H = 1.3$; he constructed 
a spatially isotropic and homogeneous
state $\omega$ which is 2nd order LTE at conformal time $\eta = \eta_1 = -1.0$,
and calculated, as just described, the ``would-be-LTE''
comparison temperatures ${\rm T}(x)
\equiv {\rm T}(\eta)$ and ${\rm T}_{\mu\nu}(x) \equiv {\rm T}_{\mu\nu}(\eta)$
numerically for earlier and later conformal times. The result is depicted in
Figure 2.
\begin{center}
\includegraphics[totalheight=5.0cm]{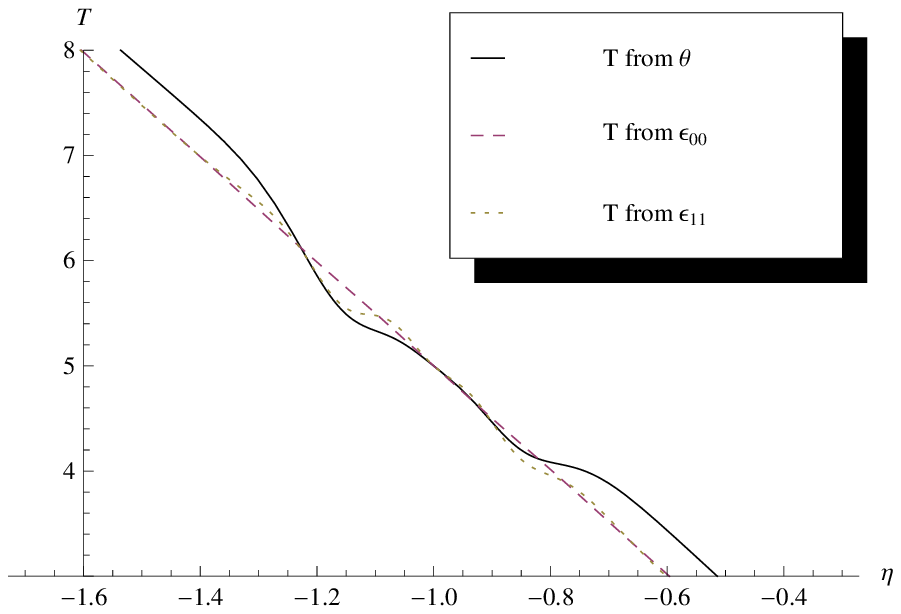} 
${}$\\
{\bf Figure 2: LTE-comparison temperatures calculated from $\theta(\eta)$ and
$\epsilon_{\mu\nu}(\eta)$ for a state constructed as 2nd order LTE state at $\eta = -1.0$ 
\cite{SchlPhD}} 
\end{center}
Figure 2 shows that the comparison temperatures deviate   from each other away from
$\eta = \eta_1 = 1.0$. However, they all drop with increasing $\eta$ as they should
for an increasing scale factor. Both the absolute and relative temperature deviations
are larger at low temperatures than at high temperatures. This can be taken as an
indication of a general effect, namely that the LTE property 
is more stable against perturbations (which drive the system away from LTE)
at high temperature than at low temperature.

\subsection{LTE and Metric Scattering}

Let us turn to a situation demonstrating that effect more clearly. We consider
the quantized, conformally coupled ($\xi = 1/6$) linear scalar field
$\Phi_{\boldsymbol{M}}$ on $\mathbb{R}^4$ with conformally flat metric of the
form (\ref{conft}), where $\Omega(\eta)$ is chosen with
\begin{align} \label{omega}
\Omega(\eta)^2 & = \lambda + \frac{\Delta \lambda}{2} ( 1 + \tanh(\rho \eta))
\end{align}
where $\lambda$, $\Delta \lambda$ and $\rho$ are real positive constants.
One can show that the quantum field $\Phi_{\boldsymbol{M}}$ has asymptotic
limits as $\eta \to \pm \infty$; in a slightly sloppy notation,
$$
   \lim_{\eta' \to \mp \infty}\, \Phi_{\boldsymbol{M}}(\eta + \eta',{\bf x}) =
   \begin{cases} \Phi_{\rm in}(\eta,{\bf x}) \\
    \Phi_{\rm out}(\eta,{\bf x})
    \end{cases} \quad (\eta \in\mathbb{R},\ {\bf x} =(x^1,x^2,x^3) \in \mathbb{R}^3)
   $$
where $\Phi_{\rm in}$ and $\Phi_{\rm out}$ are copies of the quantized linear
scalar field with mass $m$ on Minkowski spacetime, however with different
(constant) scale factors. This means that $\Phi_{\rm in}$ and $\Phi_{\rm out}$
obey the field equations
$$ (\lambda \Box +  m^2)\Phi_{\rm in} = 0\,, \quad \ \ \ ((\lambda + \Delta \lambda) \Box
 + m^2) \Phi_{\rm out} = 0 \,,$$
where $\Box = \partial_\eta^2 - \partial_{x^1}^2 - \partial_{x^2}^2 - \partial_{x^2}^2$.
If $\omega_{\rm in}$ is a state of $\Phi_{\rm in}$, it induces a state 
$\omega_{\rm out}$  of $\Phi_{\rm out}$ by setting
\begin{align*}
\langle \Phi_{\rm out}(\eta_1,{\bf x}_1) \cdots \Phi_{\rm out}(\eta_n,{\bf x}_n) & 
\rangle_{\omega_{\rm out}}
 = \langle \Phi_{\rm in}(\eta_1,{\bf x}_1)  \cdots \Phi_{\rm in}(\eta_n,{\bf x}_n) 
\rangle_{\omega_{\rm in}} \,.
\end{align*}
Thus, $\omega_{\rm out}$ is the state obtained from
$\omega_{\rm in}$ through the scattering process the quantum field $\Phi_{\boldsymbol{M}}$
undergoes by propagating on a spacetime with the expanding 
conformal scale factor as in (\ref{omega}). This particular form
of the scale factor has the advantage that the scattering transformation
taking $\Phi_{\rm in}$ into $\Phi_{\rm out}$ can be calculated
explicitly, and thus can the relation between $\omega_{\rm in}$ 
and $\omega_{\rm out}$ \cite{BirDav}. It is known that, if $\omega_{\rm in}$
is a Hadamard state, then so is $\omega_{\rm out}$.

In a forthcoming paper \cite{LinVer}, we pose the following question:
If we take $\omega_{\rm in}$ as a global thermal equilibrium state
(with respect to the time-coordinate $\eta$) of $\Phi_{\rm in}$, is
$\omega_{\rm out}$ an LTE state of 2nd order for $\Phi_{\rm out}$?
The answers, which we will present in considerably more detail in
\cite{LinVer}, can be roughly sumarized as follows.
\\[6pt]
(1) \quad For a wide range of parameters $m,\lambda,\Delta \lambda$ and
$\rho$, and inverse temperature $\beta_{\rm in}$ of $\omega_{\rm in}$,
$\omega_{\rm out}$ is a mixed temperature LTE state (with respect to
the time-direction of $\eta$) of 2nd order.
\\[6pt]
(2) \quad Numerical calculations of the comparison temperatures
${\rm T}$ and ${\rm T}_{\mu\nu}$ from $\theta$ and $\epsilon_{\mu\nu}$ for
$\omega_{\rm out}$ show that they mutually deviate (so $\omega_{\rm out}$
is no sharp temperature LTE state). Numerically one can show that the
deviations decrease with increasing temperature ${\rm T}_{\rm in}$ of
$\omega_{\rm in}$ (depending, of course, on the other parameters
$m,\lambda,\Delta \lambda$ and $\rho$).
\\[6pt]
This shows that scattering of the quantum field by the expanding
spacetime metric tends to drag the initially global thermal state 
away from equilibrium. However, this effect is the smaller the higher
the initial temperature, as one would expect intuitively. As the
initial temperature goes to zero, $\omega_{\rm in}$ approaches the
vacuum state. In this case, $\omega_{\rm out}$ is a non-vacuum state
due to quantum particle creation induced by the time-varying spacetime
metric which is non-thermal \cite{WaldScat}. That may appear surprising
in view of the close analogy of the particle-creation-by-metric-scattering
effect to the Hawking effect. However, it should be noted that the notion
of temperature in the context of the Hawking effect is different form the 
temperature definition entering the LTE condition. In the framework of 
quantum fields on de Sitter spacetime, Buchholz and Schlemmer 
\cite{BuSchl} noted that the KMS-temperature of a state with respect to
a Killing flow differs, in general, from the LTE temperature. To give a
basic example, we note that the vacuum state of a quantum field theory
in Minkowski spacetime is a KMS state at non-zero temperature with
respect to the Killing flow of the Lorentz boosts along a fixed spatial direction
(that's the main assertion of the Bisognano-Wichmann theorem, see
\cite{Haag} and references cited there). On the other hand, the vacuum state
is also an LTE state of infinite order, but at zero temperature. It appears that
the question how to modify the LTE condition such that it might become
sensitive to the Hawking temperature has not been discussed so far. 

\subsection{LTE Temperature in the Dappiaggi-Fredenhagen-Pinamonti
Cosmological Approach}

In this last part of our article, we turn to an application of the LTE concept
in an approach to cosmology which takes as its starting point the semiclassical
Einstein equation
\begin{align} \label{semiEin}
G_{\boldsymbol{M}ab}(x) & = 8\pi G \langle {\sf T}_{\boldsymbol{M}ab}(x) 
\rangle_{\omega_{\boldsymbol{M}}} \,.
\end{align}
 On the left hand side, we have the Einstein tensor of the spacetime geometry
$\boldsymbol{M}$, on the right hand side the expectation value of the stress energy
tensor of $\Phi_{\boldsymbol{M}}$ which is part of a local covariant quantum field
$\boldsymbol{M}$ in a state $\omega_{\boldsymbol{M}}$. Adopting the standing
assumptions of standard cosmology, it seems a fair ansatz to assume that a 
solution can be obtained for an $\boldsymbol{M}$ of the spatially flat 
Friedmann-Robertson-Walker  form (\ref{frw}) while taking for $\Phi_{\boldsymbol{M}}$
the quantized linear conformally coupled scalar field whose field equation
is
\begin{align}
(\nabla^a \nabla_a + \frac{1}{6} R + m^2) \Phi_{\boldsymbol{M}} = 0\,.
\end{align} 
And, in fact, in a formidable work, Pinamonti \cite{Pinam} has recently shown
that this assumption is justified: Adopting the metric ansatz (\ref{frw}) for
$\boldsymbol{M}$, he could establish that, for any given $m \ge 0$, there are
an open interval of cosmological time, a scale factor $a(t)$ and a Hadamard state
$\omega_{\boldsymbol{M}}$ such that (\ref{semiEin}) holds for the $t$-values in
the said interval (identifying $x = (t,x^1,x^2,x^3)$).
 
However, we will simplify things here considerably by setting $m = 0$
from now on, so that we have a
conformally covariant quantized linear scalar field with field equation
\begin{align}
(\nabla^a \nabla_a + \frac{1}{6} R) \Phi_{\boldsymbol{M}} & = 0 \,.
\end{align}
As pointed out above in Sec.\ 2, in the case of conformal covariant
$\Phi_{\boldsymbol{M}}$ one has to face the trace anomaly
\begin{align} \label{tracen}
 {\sf T}_{\boldsymbol{M}}{}^a{}_a(x) & = -4 Q_{\boldsymbol{M}}(x) \boldsymbol{1}
\end{align}
where $Q_{\boldsymbol{M}}$ is the divergence compensating term of (\ref{TminusQ}),
a quantity which is determined by the geometry of $\boldsymbol{M}$ and which
hence is state-independent, so (\ref{tracen}) is to be read as an equation at the
level of operators. However, equation (\ref{tracen}) is not entirely complete
as it stands, since it doesn't make explicit that ${\sf T}_{\boldsymbol{M}ab}(x)$
is subject to a renormalization ambiguity which lies in the freedom of 
adding divergence-free, local covariant tensor fields $C_{\boldsymbol{M}ab}(x)$.
Therefore, also the trace of the renormalized stress energy tensor is subject
to such an ambiguity. Let us now be a bit more specific about this point.
Suppose we define, in a first step, a renormalized $\tilde{\sf T}_{\boldsymbol{M}ab}(x)$
by the SHP renormalization prescription. Then define
$$ {\sf T}_{\boldsymbol{M}ab}(x) = {\sf T}_{\boldsymbol{M}ab}^{[0]}(x) = 
\tilde{\sf T}_{\boldsymbol{M}ab}(x) - Q_{\boldsymbol{M}}(x) g_{ab}(x)\,, $$
where $Q_{\boldsymbol{M}}(x)$ is the divergence-compensating term
defined with respect to $\tilde{\sf T}_{\boldsymbol{M}ab}(x)$. Now, if 
${\sf T}_{\boldsymbol{M}ab}(x)$ is re-defined as 
$${\sf T}^{[C]}_{\boldsymbol{M}ab}(x) = {\sf T}^{[0]}_{\boldsymbol{M}ab}(x)
  + C_{\boldsymbol{M}ab}(x)$$
  with $C_{\boldsymbol{M}ab}(x)$ local covariant and divergence-free,
one obtains for the trace
\begin{align} \label{Tr}
 {\sf T}^{[C]}_{\boldsymbol{M}}{}^a{}_a(x) & = (-4 Q_{\boldsymbol{M}}(x) 
 + C^a_{\boldsymbol{M}a}(x)) \boldsymbol{1}\,.
\end{align}
For generic choice of $C_{\boldsymbol{M}ab}$, the right-hand side of (\ref{Tr})
contains derivatives of the spacetime metric of higher than second order.
However, Dappiaggi, Fredenhagen and Pinamonti \cite{DFP} have pointed out
that one can make a specific choice of $C_{\boldsymbol{M}ab}$ such that
the right-hand side of (\ref{Tr}) contains only derivatives of the metric up to
second order.
Considering again the semiclassical Einstein equation (\ref{semiEin}) and
taking traces on both sides, one gets
\begin{align} \label{TrEin}
R_{\boldsymbol{M}}(x)  &= 8\pi G(- 4 Q_{\boldsymbol{M}}(x) + C^a_{\boldsymbol{M}a}(x))\,,
\end{align}
which (in the case of a conformally covariant $\Phi_{\boldsymbol{M}}$ considered
here) determines the spacetime geometry independent of the choice of a state
$\omega_{\boldsymbol{M}}$ (after supplying also initial conditions).
With our metric ansatz (\ref{frw}), equation (\ref{TrEin}) is equivalent to a
non-linear differential equation for $a(t)$, of the general form
\begin{align} \label{DF}
F(a^{(n)}(t),a^{(n-1)}(t),\ldots, a^{(1)}(t),a(t)) & = 0
\end{align}
with $a^{(j)}(t) = (d^j/dt^j)a(t)$ and with $n$ denoting the highest derivative
order in the differential equation for $a(t)$. For generic choices of 
$C_{\boldsymbol{M}ab}$, $n$ turns out to be greater than 2. On the other hand,
if $C_{\boldsymbol{M}ab}$ is specifically chosen so that the trace
${\sf T}^{[C]\,a}_{\boldsymbol{M}\,\ a}$ contains only derivatives up to second
order of the spacetime metric, then $n=2$. Making or not making this choice
has drastic consequences for the behaviour of solutions $a(t)$ to (\ref{DF}).
The case of generic $C_{\boldsymbol{M}ab}$, leading to $n \ge 3$ in
(\ref{DF}), was investigated in a famous article by Starobinski \cite{Star}.
He showed that, in this case, the differential equation (\ref{DF}) for $a(t)$
has solutions with an inflationary, or accelerating ($a^{(2)}(t) > 0$), phase which
is unstable at small time scales after the initial conditions. This provided
a natural argument why the inflationary phase of early cosmology would
end after a very short timespan. However,  recent astronomical observations
have shown accelerating phases of the Universe over a large timescale at late
cosmic times. The popular explanation for this phenomenon postulates an exotic
form of energy to be present in the Universe, termed ``Dark Energy'' \cite{Weib}.
In contrast, Dappiaggi, Fredenhagen and Pinamonti have shown that the specific
choice of $C_{\boldsymbol{M}ab}$ leads to a differential equation (\ref{DF})
for $a(t)$ with $n=2$ which admits stable solutions with a long-term accelerating
phase at late cosmological times \cite{DFP}. In a recent work \cite{DHMP}, it
was investigated if such solutions could account for the observations of the
recently observed accelerated cosmic expansion. Although that issue remains
so far undecided, also in view of the fact that linear quantized fields are
certainly too simple to describe the full physics of quantum effects in cosmology,
it bears the interesting possibility that ``cosmic acceleration'' and ``Dark Energy''
could actually be traced back to the renormalization ambiguities and anomalies
arising in the definition of the stress energy tensor of quantum fields in the presence
of spacetime curvature. In that light, one can take the point of view that the
renormalization ambiguity of ${\sf T}_{\boldsymbol{M}ab}$ is ultimately constrained 
(and maybe fixed) by the behaviour of solutions to the semiclassical Einstein equation:
One would be inclined to prefer such $C_{\boldsymbol{M}ab}$ which lead to
stable solutions, in spite of the argument of \cite{Star}. At any rate, only those
$C_{\boldsymbol{M}ab}$ can be considered which lead to solutions to the
semiclassical Einstein equations compatible with observational data. After all,
the renormalization freedom in quantum field theoretic models of elementary
particle physics is fixed in a very similar way.

The differential equation with stable solutions $a(t)$ derived in \cite{DFP} can be
expressed in terms of the following differential equation for the
Hubble function $H(t) = a^{(1)}(t)/a(t)$,
\begin{align} \label{Hdgl}
 \dot{H}(H^2 - H_0) = - H^4 + 2 H_0 H^2\,,
\end{align}
where $H_0$ is some universal positive constant and the dot
means differentiation with respect to $t$. There are two constant
solutions to (\ref{Hdgl}) (obviously $H(t) = 0$ is a solution) as well as
non-constant solutions depending on initial conditions. For the non-constant
solutions, one finds an asymptotic behaviour as follows \cite{DFP,Grans}:
For early cosmological times,
\begin{align} 
H(t) \approx & {} \ \, \frac{1}{t -t_0}\,, \quad a(t) \approx \, \ \Gamma (t - t_0)\,,
\end{align}
with some constants $\Gamma > 0$ and 
real $t_0$, for $t > t_0$. On the other hand, for late cosmological times:
\begin{align}
H(t) \approx & {} \ \, \sqrt{2} H_0\coth ( 2 \sqrt{2} H_0 t - 1) \,.
\end{align}
Therefore,  $H(t)$ and $a(t)$ have, for early cosmological times, a singularity
as $t \to t_0$, but different from the behaviour of a Universe filled with
classical radiation, which is known to yield \cite{WaldGR,Weib}
$$ H_{\rm rad}(t) \approx \, \ \frac{1}{2(t -t_0)}\,,  \quad  
a_{\rm rad}(t) \approx\, \ \Gamma'\sqrt{t -t_0}
 $$
with some positive constant $\Gamma'$. For the temperature behaviour of
radiation close to the singularity at $t \to t_0$ one then obtains
$$ {\rm T}_{\rm rad}(t) \approx \, \  \frac{\kappa'}{\sqrt{t - t_0}}$$
with another constant $\kappa'$. 

Now we wish to compare this to the temperature behaviour, as $t \to t_0$,
of any 2nd order, sharp temperature LTE state $\omega_{\boldsymbol{M}}$
of $\Phi_{\boldsymbol{M}}$ fulfilling the semiclassical Einstein equation
in the Dappiaggi-Fredenhagen-Pinamonti approach for $t \to t_0$. Of course,
the existence of such LTE states is an assumption, and in view of the results
of  Subsection \ref{SchlRes}, cf.\ Figure 2, this assumption is certainly an
over-idealization. On the other hand, Figure 2 can also be interpreted as saying
that, at early cosmic times, while sharp temperature LTE states might possibly
not exist, it is still meaningful to attribute an approximate temperature behaviour
to states as $t \to t_0$. 

In order to determine the temperature behaviour of the assumed 2nd order LTE state
$\omega_{\boldsymbol{M}}$, one observes first that \cite{SchlVer}
\begin{align} \label{Teps}
\langle {\sf T}_{ab}^{[C]}(x)\rangle_{\omega_{\boldsymbol{M}}} & =
\varepsilon_{ab}(x) + \frac{1}{12} \nabla^a \nabla_b \vartheta(x) +
- \frac{1}{3} g_{ab}(x) \varepsilon^c{}_c(x) \\
& \quad+ q_{\boldsymbol{M}}(x) \vartheta(x)
+ K_{\boldsymbol{M} ab}^{[C]}(x) \nonumber
\end{align}
with some state-independent tensor $K_{\boldsymbol{M} ab}^{[C]}$ which depends on $C_{\boldsymbol{M}ab}$
and which has local covariant dependence of the spacetime geometry $\boldsymbol{M}$; likewise so has
$q_{\boldsymbol{M}}$. Both $K_{\boldsymbol{M}ab}^{[C]}$ and $q_{\boldsymbol{M}}$ can be explicitly 
calculated once $C_{\boldsymbol{M}ab}$ and, consequently, $a(t)$ (respectively, $H(t)$) are specified.
Then, $\varepsilon_{ab}(x)$ and $\vartheta(x)$ are functions of $\beta(t)$ (making the
usual assumption that $\omega_{\boldsymbol{M}}$ is homogeneous and isotropic). To determine
$\beta(t)$, relation (\ref{Teps}) is plugged into the vanishing-of-divergence-equation
$$ \nabla^a \langle {\sf T}_{ab}^{[C]}(x)\rangle_{\omega_{\boldsymbol{M}}} = 0\,,$$
thus yielding a non-linear differential equation involving $\beta(t)$ and $a(t)$. Inserting
any $a(t)$ coming from the non-constant Dappiaggi-Fredenhagen-Pinamonti solutions, one can
derive the behaviour, as $t \to t_0$, of $\beta(t)$,
and it turns out \cite{Grans} that $\beta(t) \approx \gamma a(t)$ with another
constant $\gamma > 0$. This resembles the behaviour of classical radiation.
But in view of the different behaviour of $a(t)$ as compared to early cosmology of
a radiation dominated Universe, one now obtains a temperature behaviour 
$$ {\rm T}(t) \approx \, \ \frac{\kappa}{t - t_0}\,,$$
with yet another constant $\kappa > 0$;  this is more singular than
${\rm T}_{\rm rad}(t)$ in the limit $t \to t_0$.
 We will present a considerably more 
detailed analysis of the temperature behaviour of LTE states in the context of
the Dappiaggi-Fredenhagen-Pinamonti cosmological model elsewhere \cite{GrKnVe}.

\section{Summary and Outlook}

The LTE concept allows it to describe situations in quantum field theory where
states are no longer in global thermal equilibrium, but still possess, locally, thermodynamic
parameters. In curved spacetime, this concept is intriguingly interlaced with
local covariance and the renormalization ambiguities which enter into its
very definition via Wick-products and their balanced derivatives. Particularly
when considering the semiclassical Einstein equations in a cosmological context
this plays a role, and the thermodynamic properties of quantum fields in the
very early stages of cosmology can turn out to be different from what is usually
assumed in considerations based on, e.g., modelling matter as classical radiation.
The implications of these possibilities for theories of 
cosmology remain yet to be explored --- so far we have
only scratched the tip of an iceberg, or so it seems.  

There are several related developments concerning the thermodynamic behaviour
of quantum fields in curved spacetime, and in cosmological spacetimes in 
particular, which we haven't touched upon in the main body of the text. Worth
mentioning in this context is the work by Hollands and Leiler on a derivation
of the Boltzmann equation in quantum field theory \cite{HolLei}.  It should
be very interesting to try and explore relations between their approach and the
LTE concept. There are also other concepts of approximate thermal
equilibrium states \cite{DaHaP}, and again, the relation to the LTE concept
should render interesting new insights. In all, the new light that these 
recent developments shed on quantum field theory in early cosmology is
clearly conceptually fruitful and challenging.


\end{document}